\documentclass[prc,superscriptaddress,unsortedaddress,
twocolumn,showpacs,preprintnumbers,amsmath,amssymb]{revtex4}
\usepackage[dvips]{graphicx}
\usepackage{dcolumn}
\usepackage{amsmath}
\usepackage{times}
\def\la{\mathrel{\mathpalette\fun <}}

\def\fun#1#2{\lower3.6pt\vbox{\baselineskip0pt\lineskip.9pt
  \ialign{$\mathsurround=0pt#1\hfil##\hfil$\crcr#2\crcr\sim\crcr}}}

\begin{document}

\title{ Gaussian expansion approach to Coulomb breakup }

\author{T. Egami}
\email[Electronic address: ]{egami2scp@mbox.nc.kyushu-u.ac.jp}
\affiliation{Department of Physics, Kyushu University,
Fukuoka 812-8581,
Japan}
\author{K. Ogata}
\affiliation{Department of Physics, Kyushu University,
Fukuoka 812-8581,
Japan}
\author{T. Matsumoto}
\affiliation{Department of Physics, Kyushu University,
Fukuoka 812-8581,
Japan}
\author{Y. Iseri}
\affiliation{Department of Physics, Chiba-Keizai College,
Todoroki-cho
4-3-30, Inage, Chiba 263-0021, Japan}
\author{M. Kamimura}
\affiliation{Department of Physics, Kyushu University,
Fukuoka 812-8581,
Japan}
\author{M. Yahiro}
\affiliation{Department of Physics, Kyushu University,
Fukuoka 812-8581,
Japan}

\date{\today}

\begin{abstract}
An accurate treatment of Coulomb breakup reactions is presented
by using both the Gaussian expansion method and the method
of continuum discretized coupled channels.
As $L^2$-type basis functions for describing
bound- and continuum-states of a projectile,
we take complex-range Gaussian functions, which form in good approximation
a complete set in a large configuration space
being important for Coulomb-breakup processes.
Accuracy of the method is tested quantitatively for $^{8}{\rm B}+^{58}$Ni
scattering at 25.8 MeV.

\end{abstract}


\maketitle

Determination of the neutrino oscillation parameters is
one of the central issues in the neutrino physics.
The astrophysical factor $S_{17}$,
which is essentially equal to
the cross section of the $p$-capture reaction
$^7$Be($p,\gamma$)$^8$B, plays an important role
in the parameter-search procedure,
since the prediction value for the flux of $^8$B solar neutrino is
proportional to $S_{17}$. The required accuracy from astrophysics
is about 5\% in errors~\cite{Bahcall3}.
Due to difficulties of direct measurements of $^7$Be($p,\gamma$)$^8$B
at very low energies,
alternative indirect measurements were proposed;
$^8$B Coulomb
breakup~\cite{RIKEN1,RIKEN2,GSI1,GSI2,MSU1,MSU2}
is a typical example of them.

The method of continuum discretized coupled
channels~\cite{CDCC-review1,CDCC-review2} (CDCC),
which was shown to describe various projectile-breakup
processes~\cite{Yahiro1,yahiro2,sakuragi1,%
sakuragi2,sakuragi34,iseri1,Sakuragi,Surrey,Ogata},
has also been applied to $^8$B Coulomb breakup
with high degrees of
success~\cite{MSU2,Tostevin,Mortimer,Yamashita,OgataND}.
In the all analyses above, CDCC describes the reaction system
with a three-body model, i.e., $p$+$^7$Be+A as shown in Fig.~1.
\begin{figure}[htbp]
\begin{center}
 \includegraphics[width=0.3\textwidth,clip]{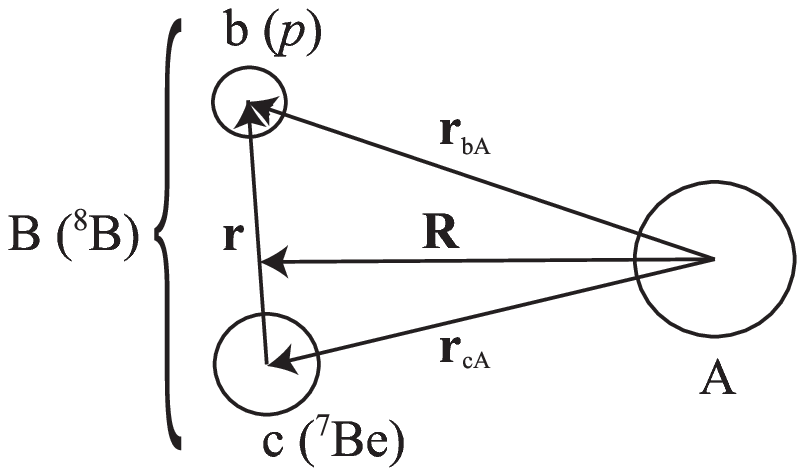}
 \caption{Illustration of a three-body (A+b+c) system.
 The symbol B=b+c stands for the projectile and A is the target.
}
\end{center}
\end{figure}

From $^8$B Coulomb breakup measured at intermediate energies,
$S_{17}$ was determined using the first-order perturbation
theory~\cite{RIKEN1,RIKEN2,GSI1,GSI2,MSU1,MSU2}.
It was shown in Refs.~\cite{Yamashita,OgataND} that
$S_{17}$ can alternatively be determined by CDCC analysis,
combined with
the asymptotic normalization coefficient
method~\cite{Xu}, of $^8$B breakup even at low energies,
such as the Notre Dame (ND) experiment~\cite{ND}.
It was found, however, that the ND data contain an
irrelevant component to $S_{17}$, corresponding to
the $1/2^-$ $^7$Be excited state in the final channel,
which of order 10\% presents in the $^8$B ground
state~\cite{Gil}; effects of this component
on $S_{17}$ can be expected around 10\%.
Therefore, the evaluated value of $S_{17}$ in
Refs.~\cite{Yamashita,OgataND}
should be assumed as the upper limit of $S_{17}$.
Even though the irrelevant component can be experimentally removed
by detecting the emitted $^7$Be in the ground state, the $1/2^-$ $^7$Be
excited state may play a dynamical role during the breakup process
of $^8$B.
Thus, description of $^8$B breakup beyond the three-body model, i.e.,
$p$+${^3}$He+${^4}$He+A four-body model,
is highly expected in order to determine $S_{17}$
with very high accuracy.

Very recently, a new treatment of breakup continuum in CDCC
was proposed~\cite{Matsumoto},
which uses pseudo-state (PS) wave functions~\cite{CDCC-review1,PS1,PS2}
obtained by diagonalizing internal Hamiltonian of the projectile
with Gaussian basis functions~\cite{H-Ka-Ki}.
For nuclear breakup processes,
CDCC with the PS method (PS-CDCC)
was found to perfectly reproduce the {\lq\lq}exact''
breakup $S$-matrix elements $S(k)$
calculated by the standard
CDCC, i.e., with the Average (Av)
method~\cite{CDCC-review1,CDCC-review2,Yahiro1,Piya,YK1}
for discretization of breakup continuum.
The most important feature of PS-CDCC is that it
generates $S(k)$ as a smooth function of $k$ in its entire range,
without assuming any form {\it a priori} for $k$-distributions,
in contrast to CDCC with the Av method (Av-CDCC).

Additionally, as discussed in Refs.~\cite{Matsumoto,H-Ka-Ki},
PS-CDCC with Gaussian basis functions makes
it possible the four-body CDCC analysis of projectile
breakup, a preliminary result of which
for $^6$He elastic scattering,
including effects of breakup channels of $^6$He,
has already been obtained~\cite{Matsumoto2}.
Before making four-body PS-CDCC analysis of $^8$B Coulomb breakup,
however, one must see the applicability of PS-CDCC to breakup
processes due to long-ranged Coulomb coupling potentials.
This is quite nontrivial because the $r$-space of the projectile
needed to describe Coulomb breakup is much larger than that in
the case of nuclear breakup;
for example, the effective space is $r \la 20$ fm for nuclear breakup
of $d$ or $^{6}$Li~\cite{Matsumoto},
but $r \la 100$ fm for Coulomb breakup
of $^{8}$B~\cite{Yamashita,OgataND}.

The aim of this brief report is,
as the first step towards the four-body CDCC analysis of $^8$B
Coulomb breakup, to show that three-body PS-CDCC based on the
Gaussian basis functions can well describe the corresponding
result of three-body Av-CDCC for $^8$B breakup.
As for the test case we take $^8$B breakup from $^{58}$Ni at
25.8 MeV.

Below we briefly recapitulate the formulation of three-body CDCC.
We consider the three-body system of Fig.~1.
The model Hamiltonian of the system is
\begin{eqnarray}
H&=&K_{r}+V_{\rm bc}({\bf r})+ K_{R} +
U_{\rm bA}({\bf r}_{\rm bA})+ U_{\rm cA}({\bf r}_{\rm cA}) \; .
\nonumber
\end{eqnarray}
Coordinates are defined in Fig.~1.
Operators $K_{r}$ and $K_{R}$ are kinetic energies
associated with ${\bf r}$ and ${\bf R}$, respectively, and
$V_{\rm bc}({\bf r})$ is the interaction between b and c.
The interaction
$U_{\rm bA}$ ($U_{\rm cA}$) between b (c) and A
is taken to be
the optical potential for b+A (c+A) scattering;
Coulomb breakup is induced by the Coulomb components of
the potentials.
For simplicity, in this study we neglect the intrinsic spins of
individual constituents of the system.

In CDCC,
the states of the projectile are classified by the linear and the
angular momenta, $k$ and $\ell$,
of relative motion between b and c,
which are truncated by
$k \leq k_{\rm max}$ and $ \ell \leq \ell_{\rm max}$.
The truncation is the most basic assumption in CDCC, and
it is confirmed that
calculated $S$-matrix elements converge
for sufficiently large $k_{\rm max}$ and $\ell_{\rm max}$%
~\cite{CDCC-review1,Yahiro1,Piya}. The converged
CDCC solution is the unperturbed solution of the distorted
Faddeev equations,
and corrections to the solution are
negligible within the region of space in which the reaction
takes place~\cite{CDCC-foundation}.
The $k$-continuum of the b+c system are then discretized into a finite
number of states,
each of which corresponds to a {\lq\lq}discretized-continuum state''
with a certain positive eigenenergy labeled by $i$.
The resulting orthonormalized wave functions,
$\{ \hat{\Phi}_{i\ell}(r)i^\ell Y_{\ell m}(\Omega_r); \,
i=1 \mbox{--}N \}$,
are assumed to form an approximate complete set in
the $r$- and $k$-space being important for the breakup reaction.

The three-body wave function $\Psi$ of the system is expanded by
$\{ \hat{\Phi}_{i\ell}(r)i^\ell Y_{\ell m}(\Omega_r)\}$
and then inserted into the approximate
three-body Schr\"{o}dinger equation
$(H-E)\Psi=0$. One can then obtain
a set of coupled differential equations,
called CDCC equations, that
provide the discrete $S$-matrix element,
$\hat{S}_{\gamma}$,
for the transition from the initial channel
to a discretized-continuum one $\gamma=(i,\ell)$.
In order to calculate a breakup cross section of the projectile
with a certain range of breakup energies, or a
coincidence-cross-section, a smoothing procedure, which
constructs continuous $S_\ell (k)$ from the discrete
$\hat{S}_{\gamma}$, is necessary.
Actually, this is the case in the analysis of $^8$B Coulomb-breakup
experiments~\cite{RIKEN1,RIKEN2,GSI1,GSI2,MSU1,MSU2,ND}.

Discretization of the breakup continuum
in the PS method is done by
diagonalizing the internal Hamiltonian
$H_{\rm bc}=K_{r}+V_{\rm bc}({\bf r})$
in a space spanned by a finite number of
$L^2$-type basis-functions, for which we here take
the following pairs of functions~\cite{H-Ka-Ki},
\begin{eqnarray}
\phi^{\rm C}_{j\ell}(r)
&=&
r^{\ell}\exp\left[-(r/a_{j})^2\right]
\cos\left[\,b\,(r/a_{j})^2\right],
\nonumber \\
\phi^{\rm S}_{j\ell}(r)
&=&
r^{\ell}\exp\left[-(r/a_{j})^2\right]
\sin\left[\,b\,(r/a_{j})^2\right],
\quad (j = 1 \mbox{--} n),
\nonumber
\label{eq:comp-g}
\end{eqnarray}
where $\{a_{j}\}$ are
assumed to increase in a geometric
progression and $b =\pi/2$.
We refer to the basis as the complex-range Gaussian basis,
since the basis functions
can be expressed by Gaussian functions with a complex-range
parameter,
$r^{\ell}\exp[-(1+i b)(r/a_j)^2]$, and its complex conjugate.
The complex-range Gaussian basis functions
are oscillating with $r$, so they can simulate
the oscillating pattern of the continuous breakup-state wave-functions
as shown in Ref.~\cite{H-Ka-Ki}, which is very important for
the description of Coulomb breakup by PS-CDCC.
An accurate transformation from $\hat{S}_{\gamma}$
to $S_\ell (k)$:
\begin{equation}
   S_{\ell}(k) \approx \sum_{i} f_{i\ell}(k)
    \hat{S}_{\gamma} \; ,
\label{S-matrix-appro}
\end{equation}
with
$f_{i\ell}(k)=\langle\Phi_{\ell}(k,r)|\hat{\Phi}_{i\ell}(r)\rangle$,
is possible
when the basis functions form an approximate complete set in
the finite configuration space being important for the breakup
process~\cite{Matsumoto}.

In the Av method, on the other hand,
the $k$-continuum [0, $k_{\rm max}$], for each $\ell$,
is divided into a finite number of bins,
each with a width $\Delta_{i\ell}=k_{i}-k_{i-1}$, and
the continuum breakup-states in the $i$ th bin
are averaged with a weight function
$w_{i\ell}(k)$~\cite{CDCC-review1,CDCC-review2};
we in this study take $w_{i\ell}(k)=1$ for simplicity.
The resulting orthonormal state is then described as
\begin{equation}
\label{state-AV}
 \hat{\Phi}_{i\ell}(r) =
 \frac{1}{\sqrt{\Delta_{i\ell}}} \int_{k_{i-1}}^{k_{i}}
\Phi_{\ell}(k, r) dk
\quad
 {\mbox{(for Av).}}
\end{equation}
Inserting Eq.~(\ref{state-AV}) into Eq.~(\ref{S-matrix-appro})
leads to
$S_{\ell}^{\rm Av}(k)
=\hat{S}_{\gamma,\gamma_0}/\sqrt{\Delta_{i\ell}}$
for $k_{i-1} < k \leq k_i$,
as shown in Ref.~\cite{Tostevin}.

It should be noted that
$S_{\ell}^{\rm Av}(k)$ is
smooth only within each bin,
while $S_{\ell}(k)$ with the PS method, i.e., $S_{\ell}^{\rm PS}(k)$,
is a smooth function of $k$ in its entire range.
Details of the discretization and smoothing procedures
both for the Av and PS methods can be found in Ref.~\cite{Matsumoto}.

In order to see the applicability of PS-CDCC for Coulomb breakup,
we calculated the breakup cross section of $^8$B+$^{58}$Ni
scattering at 25.8 MeV.
In this process both nuclear and Coulomb interactions, with
interference, play important roles. Moreover,
first order perturbation theory cannot reproduce the experimental
data~\cite{ND} at all, since higher-order processes cannot be neglected
at such a low incident energy~\cite{EB2}.
Thus, it is a good test for PS-CDCC to compare the results with
those calculated by Av-CDCC that reproduce the experimental data
very well~\cite{Tostevin,Yamashita,OgataND}.

For both the Av and PS methods, we took the $^8$B single-particle
wave-function of Esbensen and Bertsch~\cite{EB} and the same potential
as for the ground state was used to generate wave functions in
scattering states;
the depth of the potential is fitted to reproduce the separation
energy of the proton, i.e., 137 keV.
We included only s- and p-states of $^8$B for
saving computation time.
It was shown in
Refs.~\cite{Tostevin,Yamashita,OgataND} that d- and f-states are
necessary to reproduce the experimental
data~\cite{ND}.
Our aim in this
brief report, however, is to see the convergence between two
theoretical calculations.
As for the distorting potential for $p$-$^{58}$Ni ($^7$Be-$^{58}$Ni),
we adopted the parameter set of Yamashita~\cite{Yamashita}
(Moroz {\it et al.}~\cite{Moroz}).
The maximum total-angular-momentum $J_{\rm max}$ is 1000 and
$R_{\rm max}$ is taken to be 500 fm. CDCC equations were
solved with the predictor-corrector Numerov method with
stabilization~\cite{PCN}.

%
\begin{figure}[htbp]
 \begin{center}
  \includegraphics[width=0.3\textwidth,clip]{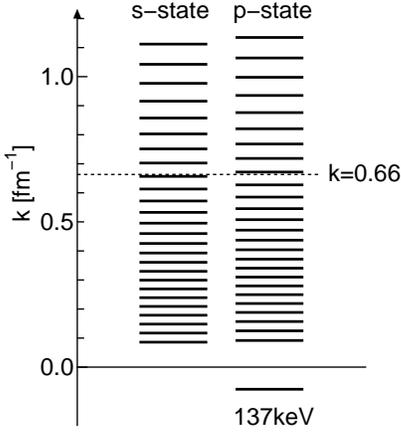}
\caption{Discretized momenta for $^8$B; the left (right) side corresponds
to the s-state (p-state).
The horizontal dotted line represents the cutoff
momentum $k_{\rm max}$ taken to be 0.66 fm$^{-1}$.}
 \end{center}
\end{figure}
In the Av method, the modelspace with $k_{\rm max}=0.66$ fm$^{-1}$ and
$\Delta_{i\ell}=0.66/16$ (0.66/32) fm$^{-1}$ for $\ell=1$ (0)
gives convergence of the resulting total breakup cross section.
The maximum internal coordinate $r_{\rm max}$ was taken
to be 100 fm. In order to obtain the correct asymptotic form
of the Coulomb coupling potentials, we made the following
approximation for the monopole components:
\begin{eqnarray}
  v^{\rm Coul}_{i'\ell' i\ell; \lambda=0}(R)
  &=&
     \int_0^{r_{\rm max}}
     \hat {\Phi}_{i'\ell'}^* (r)
     \frac{Z_{\rm j}Z_{^{58}{\rm Ni}}e^2}{R}
     \hat {\Phi}_{i\ell} (r)
     r^2 dr
  \nonumber \\
  &\approx&
     \frac{Z_{\rm j}Z_{^{58}{\rm Ni}}e^2}{R}
     \delta_{i'i}\delta_{\ell'\ell},
\label{screen}
\end{eqnarray}
where ${\rm j}=p$ or $^7$Be. In the practical calculation
we took account for Coulomb radii for $p$-$^{58}$Ni and
$^7$Be-$^{58}$Ni and Eq.~(\ref{screen}) has a slightly
complicated form~\cite{Yamashita}.
It should be noted that Eq.~(\ref{screen})
is exact when $r_{\rm max}\rightarrow \infty$;
the numerical results shown below were found to converge
at $r_{\rm max}=100$ fm.
The result thus obtained, i.e., with Av-CDCC, is assumed to be
the {\lq\lq}exact'' solution.
%
\begin{figure}[bp]
 \begin{center}
  \includegraphics[width=0.35\textwidth,clip]{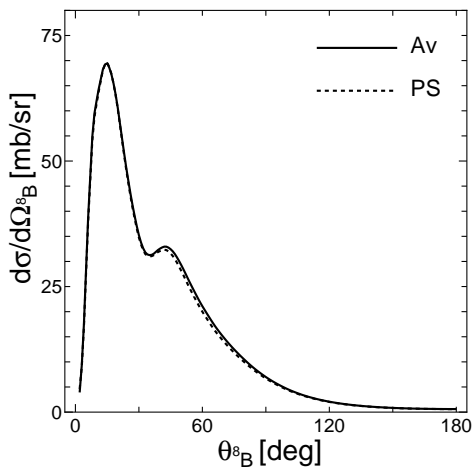}
  \caption{Angular distribution of the total breakup cross section
  for $^{58}$Ni($^8$B,$^8$B$^*$) at 25.8 MeV. The solid and dashed
  lines represent the results with the Av and PS methods,
  respectively.}
 \end{center}
\end{figure}

In the PS method, we used the complex-range Gaussian basis
with ($a_1=1.0$, $a_{n}=35.0$, $2n=60$, $b=\pi/2$) that
gives good convergence. The number of channels included in
the CDCC calculation, i.e., $k \le k_{\rm max} = 0.66$ fm$^{-1}$,
was 18 for both the s- and p-states. The resulting wave functions
with positive eigenenergies turned out to oscillate up to
about 100 fm.
In order to obtain orthonormality of the
$\hat{\Phi}_{i\ell} ({\bf r})$, we put $r_{\rm max}=130$ fm
in the calculation of coupling potentials.
We show the level sequences of the resulting discrete eigenstates
in Fig.~2. One sees each interval of the levels is almost even, which
is just the same as in Ref.~\cite{Matsumoto}.

We show in Fig.~3 the calculated angular distribution of $^8$B
total breakup cross section for $^{58}$Ni($^8$B,$^8$B$^*$)
at 25.8 MeV. The solid and dashed lines correspond to
the Av and PS methods, respectively. One sees both results
coincide with very high accuracy,
which implies that a forthcoming four-body PS-CDCC analysis
of $^8$B Coulomb breakup has enough accuracy
to determine a reliable value of $S_{17}$, i.e., within 5\% of errors.

%
\begin{figure}[bp]
 \begin{center}
  \includegraphics[width=0.47\textwidth,clip]{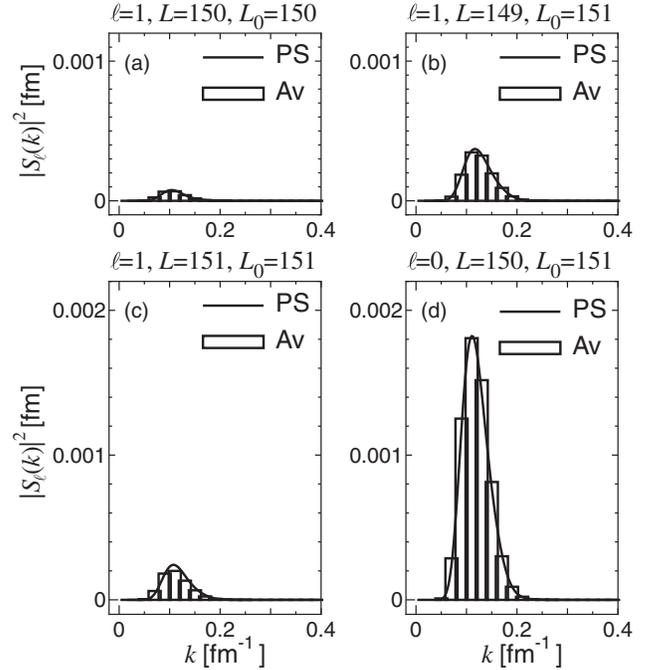}
  \caption{The squared moduli of breakup $S$-matrix elements,
  as a function of $k$, at $J=150$ for $^8$B+$^{58}$Ni
  scattering at 25.8 MeV.
  The panel (a), (b), (c) and (d) correspond to
  $(\ell,L,L_0)=(1,150,150)$, (1,149,151),
  (1,151,151) and (0,150,151), respectively.
  In each panel, the solid line represents the result of PS-CDCC,
  while the step line is the result of Av-CDCC assumed
  as the {\lq\lq}exact'' $S$-matrix elements.}
 \end{center}
\end{figure}
In order to see the validity of PS-CDCC for the $^8$B Coulomb
breakup more strictly, below
we compare $S_\ell^{\rm PS}(k)$ with $S_\ell^{\rm Av}(k)$.
In the calculation of the latter, i.e., the {\lq\lq}exact'' $S_\ell (k)$,
we put $\Delta_{i\ell}$ for the p-state to be 0.66/32 fm$^{-1}$.
This makes the $k$-dependence of $S_\ell^{\rm Av}(k)$
clearer, since $S_\ell^{\rm Av}(k)$ is constant in $k$-region
corresponding to each bin with the width of $\Delta_{i\ell}$,
as shown in Ref.~\cite{Matsumoto}.
It should be noted that this refinement of
the modelspace for the Av method makes no changes in the physical
quantities such as the elastic and total breakup cross sections.

Figure 4 shows the result of the comparison of
$|S_\ell (k)|^2$ at $J=150$, which
corresponds to the scattering angle of 10$^\circ$ assuming
the classical path. It was found that CDCC calculation with only
Coulomb coupling potentials gives a peak at 10$^\circ$ in the
total breakup cross section. Thus, it can be assumed that Fig.~4
corresponds to the most-Coulomb-like breakup process;
in any case, features of the result were found to be
almost independent of $J$.
In each panel of Fig.~4 the result with ($\ell,L,L_0$),
where $L$ ($L_0$) is the orbital angular momentum between $^8$B and
$^{58}$Ni in the final (initial) channel, is shown.
The panel (a), (b), (c) and (d) correspond to
$(\ell,L,L_0)=(1,150,150)$, (1,149,151), (1,151,151)
and (0,150,151), respectively;
all other components are negligibly small and not shown in
the figure. One sees that the result of PS-CDCC (solid line)
very well reproduces the {\lq\lq}exact'' solution (step line)
for all $k$ being significant for the $^8$B {\lq\lq}Coulomb''
breakup.

In summary, PS-CDCC proposed in Ref.~\cite{Matsumoto} is shown to
describe Coulomb breakup processes very well.
Due to the long-ranged Coulomb coupling-potentials, the modelspace
required for CDCC is very large. Particularly, one must prepare
the internal wave functions of the projectile, both in bound and
continuum states, for a wide range of the internal coordinate,
say, 0--100 fm, which is in general difficult for PS methods.
We found that this can easily be achieved by using the complex-range
Gaussian basis, in the case of two-body projectile.
According to Ref.~\cite{H-Ka-Ki}, the basis is also
applicable to the reaction processes with three- and four-body
projectiles, since energies and wave functions of the pseudo-states
of the projectiles are given easily.
Moreover, all coupled-channel potentials in four- and five-body PS-CDCC
can analytically be given by expanding each nuclear-optical-potential
concerned in terms of Gaussian functions.
Then, PS-CDCC based on the complex-range Gaussian basis
functions is expected to become an effective method
of practical use even for four-body Coulomb
breakup of $^{8}$B, in which $^{8}$B is assumed to
be a bound state of the $p$+${^3}$He+${^4}$He system.
Such a four-body analysis
of $^{8}$B dissociation is desirable in order to determine $S_{17}$
with high accuracy and then approach to the solar neutrino problem from
the nuclear physics side.

\section*{Acknowledgments}
The authors would like to thank M. Kawai and T. Motobayashi
for helpful discussions.
This work has been supported in part by the Grants-in-Aid for
Scientific Research (12047233, 14540271)
of Monbukagakusyou of Japan.


\end{document}